\documentclass[12pt]{article}
\usepackage{latexsym,graphicx,multirow}
\usepackage{float}
\usepackage{amssymb}
\usepackage{amsmath}
\usepackage{amscd}
\usepackage{amsthm}
\usepackage[left=2cm,top=2.5cm,right=2.5cm,bottom=1.5cm]{geometry}
\usepackage{hyperref}
\usepackage{epstopdf}
\usepackage{cite}
\usepackage{caption}
\usepackage{subcaption}
\usepackage{color}

\begin{document}
	
\begin{center}
\large{\bf{Barrow entropic Quintessence and Dilation dark energy Models with Generalized HDE cut-off}} \\
		\vspace{10mm}
\normalsize{Priyanka Garg$^1$, Vinod Kumar Bhardwaj$^2$, Anirudh Pradhan$^3$  }\\
		\vspace{5mm}
\normalsize{$^{1,2}$Department of Mathematics, Institute of Applied Sciences and Humanities, GLA University, Mathura-281 406, Uttar Pradesh, India}\\
\vspace{2mm}
\normalsize{$^{3}$Centre for Cosmology, Astrophysics and Space Science (CCASS), GLA University,\\	Mathura -281 406, Uttar Pradesh, India}\\
			\vspace{2mm}
$^2$E-mail:dr.vinodbhardwaj@gmail.com\\
	\vspace{2mm}
	$^1$E-mail:pri.19aug@gmail.com\\
		\vspace{2mm}
$^3$E-mail:pradhan.anirudh@gmail.com \\
			\vspace{10mm}
		
\end{center}
	
\begin{abstract}

In the present work, we have analyzed the behaviors of extension of generalized Barrow holographic dark energy(`BHDE'). A ``generalized BHDE model based on the particle and the future horizon using infrared cut-off" was proposed by Nojiri et al. (2022). In this work, we have reviewed the generalized BHDE extension under the assumption of a generalized HDE cut-off.  Using a scale factor of the form $a = k t^m$, the dynamics of the cosmos have been discussed through graphic demonstration. By applying the ``open-source emcee Python package", the values of the free parameters $k$ and $m$ are estimated on 57 OHD points by the Markov Chain Monte Carlo (MCMC) technique. We have examined the behavior of the equation of state (EoS) parameter, $( p_{de})$, and dark energy density $(\rho_{de})$. We have also discussed the equivalence of holographic dark energy (DE) with the Barrow entropic DE and its extension.  Also, we have explained quintessence and dilation dark energy models in the context of Barrow entropic DE.
\end{abstract}
	

	
\section{Introduction}
During last two decades, a number of observations, including type Ia supernovae, CMB radiations, large scale structure (LSS), the Sloan Digital Sky Survey (SDSS), the Wilkinson Microwave Anisotropy Probe (WMAP), and Planck observations \cite{ref1,ref2,ref3,ref4,ref5,ref6}, have suggested that our universe is expanding with acceleration, this is due to some unknown exotic fluid  known as dark energy (DE). The Cosmological Constant ( $\Lambda$ ) is assumed as most effective alternate of DE to explain the accelerated expansion of the universe\cite{ref5}. Several models have been suggested to explain the nature of the cosmological constant\cite{ref7,ref8,ref9,ref10,ref11}.\\

Due to its direct connection with `space-time', the `holographic dark energy (HDE)' has much consideration. The vacuum energy's cosmic behavior is made clear by HDE. The existing cosmic acceleration does not found in the `HDE models' with `Hubble' radius as `IR cut-off', although it seen in the models having event horizon as cut-off \cite{ref12}. Akhlaghi \cite{ref13} described the HDE models with Granda-Oliver, Ricci scale, and future horizon cut-offs to explain the evaluation and accelerated growth of the universe. Holographic dark energy model with Granda-Oliver cut-off have been examined by Ghaffari \cite{ref14}. If the mass of black hole is greater than the vacuum energy, ``the horizon length L is considered as IR cutoff" in black hole thermodynamics \cite{ref15}. In cosmology, the holographic principle is generally adopted to describe the dark energy(DE) epoch \cite{ref15a}.\\

BHDE is one of the alternative forms of dark energy, which is based on the newly suggested Barrow entropy instead of the standard Bekenstein-Hawking (BH) entropy \cite{ref16,ref17,ref18,ref19}. ``Saridakis et al.\cite{ref20} have studied the generalized second rule of thermodynamics using the Barrow entropy on the horizon". Mamon et al. \cite{ref21} have investigated the validity of ``BHDE models" by taking the dynamical apparent horizon into account as the thermodynamic boundary. The Barrow entropy is also used by Saridakis \cite{ref22} to present the modified cosmic model. The compatibility of the ``BHDE models" with observational data has been shown by Anagnostopoulos et al. \cite{ref23}. Various researchers \cite{ref24,ref25,ref26,ref27,ref28,ref29,ref30,ref31,ref33,ref34,ref34a,ref34b} have studied BHDE models in different contexts.\\

Since HDE models are based on `holographic principle' instead of introducing a term in Lagrangian, they differ significantly from conventional DE models. Nojiri et al.\cite{ref35} claimed that the BHDE model is equivalent to the generalized HDE model. The proposed generalized BHDE model depends on future and particle horizons by taking IR as a cut-off. In the direction of generalized entropies, few remarkable studies can be seen in refs. \cite{ref35aa,ref35bb,ref35cc,ref35dd}. Inspired by this, the authors in the present manuscript describe an extension of generalized BHDE by assuming a Generalized HDE cut-off. 
\\

In this study, the authors analyzed the extension of Generalized BHDE by assuming generalized HDE cut-off. By considering the power law $a = k t^m$, the cosmos dynamics have been discussed by graphical depiction. Applying the ``open-source emcee Python package", the model's free parameters are estimated with 57 OHD points utilizing the ``MCMC technique". The present study is organized as: In section $2$ we have presented the Thermodynamics of space-time. We proposed the solution of the field equations with Generalized HDE-cutoff in Section $3$. In section $4$, we have explained the equivalence of generalized HDE with the extension of barrow entropic DE. In Section $5$, we have discussed the power law cosmology. In Section $6$, the methodology for estimation of the model's free parameters on the latest 57 OHD data points has been discussed. In Section $7$, we discuss with Quintessence field model. We have explained the dilation field in Section $8$. The concluding remarks are mentioned in Section $9$.

\section{Thermodynamics of space-time and cosmology}
Gravity thermodynamics is typically described by the ``Bekenstein Hawking (BH) area law $S_{BH} = A/(4G) $". It is applicable to both the apparent horizon of the universe and the entropy of black-hole horizons. On the basis of non-extensive generalizations of the statistics of the horizon degrees of freedom or quantum gravitational deformations of the horizon geometry, a number of changes to entropy have been suggested. Tsallis \cite{ref35a}, and Kaniadakis \cite{ref35b} entropies are two particular instances among them.  The use of such entropies can be seen in \cite{ref35c,ref35d,ref35e,ref35f,ref35g}.  Barrow \cite{ref16} has developed a new generalized entropy based on a modified horizon supplied with a fractal structure.
	
As ``Barrow entropy" was developed for ``black holes" but it can be used in a cosmic context according to the gravity-thermodynamic conjecture. In this approach, the Barrow entropy-driven corrections to the Friedmann equations in the Standard Model of Cosmology (SMC) are obtained. Additionally, the holographic principle can be used in conjunction with Barrow entropy to produce Barrow holographic dark energy \cite{ref20,ref22,ref30}. As a result, one can apply observational data to the aforementioned structures to derive restrictions on the Barrow exponent $\Delta$ \cite{ref23,ref41}.  All of these investigates find that variations from the BH entropy as predicted are relatively small.

The black hole entropy is expressed as
\begin{equation}
\label{1}
S = A/4G,~~~~ A = 4 \pi {r^2_H} 
\end{equation}\\
Here, $S$ stands for the Bekenstein-Hawking (BH) entropy and $r_{H}$ represents the horizon's radius. The relationship between gravity and thermodynamics is defined in a lot of recent studies \cite{ref38,ref39}. The first law of thermodynamics may also be defined using the FLRW equations when the BH entropy and apparent horizon are considered as the ``thermodynamics of space-time". ``Barrow recently asserted that quantum-gravitational processes, which are inspired by the Covid-19 viral pictures, might be used to introduce the fractal and complicated aspects to the black-hole structure". The ``Barrow entropy" is read as \cite{ref19}: 
\begin{equation}
\label{2}
S =  \frac{A_0}{4G} \left( \frac{A}{A_0} \right)^{1+ \Delta}, 
\end{equation}
where, $A_0$ is a constant. For $\Delta = 0$, a quantum gravitational deformation exists and most fractal black hole structure is obtain for $\Delta = 1$. When the Barrow entropy is applied to cosmology, the Friedmann equations also transformed, and these transformations could be seen as a source of dark energy density \cite{ref24,ref25,ref26,ref27,ref41,ref42}.

\section{Barrow Entropy with Generalized HDE cut-off }
We consider the ``flat FLRW space-time metric" as:
\begin{equation}
\label{3}
ds^{2}=   -dt^{2} + a^2 (t) \left[ (dx^{1})^2 + (dx^{2})^2 + (dx^{3})^2 \right] ~,
\end{equation}
here, the scale factor $a(t)$ is the function of time.\\
The cosmic horizon radius is defined as:
\begin{equation}
\label{4}
r_H =\frac{1}{\left( \alpha H^2 + \beta \dot H  \right)^\frac{1}{2}} 
\end{equation}
where, $H = \frac{\dot a}{a}$ is the Hubble parameter. \\
The change in heat can be given by
\begin{eqnarray}
\label{5}
dQ & = & -dE = - \frac{4 }{3} \pi {\dot \rho} {{r_H^3}}  dt \nonumber\\
& = & - \frac{4}{3} \pi {\left( \alpha H^2 + \beta \dot H  \right)^\frac{-3}{2}} {\dot \rho} dt \nonumber\\
& = & {{4 \pi}} \left({\rho +p}\right) {{\left( \alpha H^2 + \beta \dot H  \right)^\frac{-3}{2}}}  H dt
\end{eqnarray}

Utilizing first law of thermodynamics   $T dS = dQ$ and law of conservation $\dot{\rho}+3\rho H +3 p H = 0$, we get
\begin{equation}
\label{6}
T \frac{dS}{dt} = \frac{{4 \pi} \left({\rho + p}\right)}{\left( \alpha H^2 + \beta \dot H  \right)^\frac{3}{2}}  H  
\end{equation}
The above expression along with the Hawking temperature defined by \cite{ref40},
\begin{equation}
\label{7}
T = \frac{1}{2 \pi r_H} = \frac{{\left( \alpha H^2 + \beta \dot H  \right)^\frac{1}{2}}}{2 \pi}
\end{equation}
Second FLRW equation as 
\begin{equation}
\label{8}
\dot H = - 4 \pi G \rho \left(1 + \frac{p}{\rho} \right)
\end{equation}
which on integrating provides the first FLRW equation
\begin{equation}
\label{9}
H^2 =  \frac{1}{3} 8 \pi G \rho + \frac{1}{3} \Lambda
\end{equation}
here, the integration constant is $\Lambda $, which is considered as a cosmological constant.\\
Similarly, for the Barrow entropy using Eqs. (8), (9), and Eq. (2), we get the expression
\begin{equation}
\label{10}
\frac{dS}{dt} = \frac{dS}{dA} \frac{dA}{dt}
\end{equation}
{since}
\begin{eqnarray}\label{11}
\frac{dA}{dt} &=& -4 \pi \left( \alpha H^2 + \beta \dot H  \right)^{-2} (2 \alpha H \dot H + \beta \ddot H)\nonumber\\
\frac{dS}{dt} &=& - {4 \pi} \left(\frac{1+ \Delta}{4 G}\right) \frac{(2 \alpha H \dot H + \beta \ddot H)}{\left( \alpha H^2 + \beta \dot H  \right)^{2}} \left(\frac{{{H_1}^2}}{ \alpha H^2 + \beta \dot H} \right)^{\Delta} 
\end{eqnarray}
where $A_0 = \frac{4 \pi}{{H_1}^2}$, ${H_1}$  is constant.\\
The second FLRW equation for the Barrow entropy is obtained as
\begin{equation}
\label{12}
\frac{({1+ \Delta})}{H} (2 \alpha H \dot H + \beta \ddot H) \left(\frac{{{H_1}^2}}{ \alpha H^2 + \beta \dot H} \right)^{\Delta}    = - 4 \pi G (\rho + p)
\end{equation}
On integrating the above equation, we get,
\begin{equation}
\label{13}
\frac{({1+ \Delta})}{{(1- \Delta)}} {H_1}^2  \left(\frac{{{H_1}^2}}{ \alpha H^2 + \beta \dot H} \right)^{\Delta - 1}    = 
\frac{8 \pi }{3} G \rho + \frac{1}{3} \Lambda
\end{equation}
Now, FLRW Eqs. (8) and (9) can be transformed into,
\begin{equation}
\label{14}
\dot{H} = - 4 \pi G ~ \left[(\rho_B + \rho) + \left(p_B + p\right)\right]
\end{equation}

\begin{equation}
\label{15}
H^{2}=\frac{8 \pi G}{3}\left(\rho_{\mathrm{B}} +\rho   \right) + \frac{\Lambda}{3}
\end{equation} 

For the Barrow entropy,  from Eqs. (12)-(15), the effective energy density $\rho_B$ and pressure $p_B$ are expressed as   
\begin{equation}
\label{16}
\rho_B= \frac{3}{8 \pi G} \bigg[H^2 - \frac{({1+ \Delta})}{{(1- \Delta)}}
{H_1}^2  \left(\frac{{{H_1}^2}}{ \alpha H^2 + \beta \dot H} \right)^{\Delta - 1} \bigg]
\end{equation} 
\begin{eqnarray} 
\label{17}
p_B &=&\frac{1}{4 \pi G}\bigg[\frac{({1+ \Delta})}{H} (2 \alpha H \dot H + \beta \ddot H) \left(\frac{{{H_1}^2}}{ \alpha H^2 + \beta \dot H} \right)^{\Delta} - \dot H\bigg]\nonumber\\
&-& \frac{3}{8 \pi G} \bigg[H^2 - \frac{({1+ \Delta})}{{(1- \Delta)}}
{H_1}^2 \left(\frac{{{H_1}^2}}{ \alpha H^2 + \beta \dot H} \right)^{\Delta - 1} \bigg] 
\end{eqnarray}

The ``EoS parameter for the Barrow entropy" can be expressed as 
\begin{eqnarray}
\label{18}
\omega_B &=&\frac{2 \left[ \frac{({1+ \Delta})}{H} \left(\frac{{{H_1}^2}}{ \alpha H^2 + \beta \dot H} \right)^{\Delta} (2 \alpha H \dot H + \beta \ddot H) - \dot H   \right] }{ 3 \bigg[H^2 - \frac{({1+ \Delta})}{{(1- \Delta)}}
	{H_1}^2  \left(\frac{{{H_1}^2}}{ \alpha H^2 + \beta \dot H} \right)^{\Delta - 1}\bigg]}-1
\end{eqnarray}
The EoS parameter $\omega_B$ for the energy density of Barrow entropy follows the ``$\dot \rho_B + 3 H \rho_B (1 + \omega_B) = 0$". Moreover, Eq. (18), explain that the EoS parameter tremendously depend on the exponent $\Delta$. For the different values of exponent $\Delta$, BHDE can exist in a quintessence region, in a phantom era, or may cross the phantom-divide during the cosmic evolution \cite{ref22}. 
\section{Equivalence of generalized ``Holographic Dark Energy with the Extension of Barrow entropic Dark Energy"}
We have study the models where the entropy exponent shows an extending behavior particularly, when the universe is expanding. The entropic dark energy models with variable exponent has been discussed in \cite{ref43,ref44}. Here, the authors claimed that the behavior in this case is caused by a physical degree of freedom that corresponds to entropy. The renormalization of a quantum theory also implies that the degrees of freedom depend on the scale. We express a dimensionless variable in cosmology $x=H_{1}^{2} / H^{2}$, where $H_{1}^{2}=4 \pi / A_{0}$, as the Hubble parameter determines the energy scale. On applying this expanded formalism to the Barrow entropy (where the exponent of each entropy function varies), then the Barrow entropy function can be recasts as, 

\begin{equation}
\label{19}
S_{B}= \left(\frac{A}{A_{0}}\right)^{1+\Delta(x)} A_{0} \frac{1}{4 G} ~ .
\end{equation}

Using $A=4 \pi r_{h}^{2}$, We deduce from equation (19)

\begin{equation}
\label{20}
\frac{d S_{B}}{d t}=\frac{\partial S}{\partial A} \frac{d A}{d t}+\frac{\partial S}{\partial x} \frac{d x}{d t}
\end{equation}
\[
\frac{dS_B}{dt}  = 
-\frac{1}{4 G}\left(\frac{4 \pi (2 \alpha H \dot H + \beta \ddot H )}
{(\alpha H^2 + \beta \dot H)^{2}}\right)\left(\frac{H_{1}^{2}}{{(\alpha H^2 + \beta \dot H)}}\right)^{\Delta(x)}
\] 
\begin{equation}
\label{21}
\left\{(1+\Delta(x))+\frac{H_{1}^{2}}{(\alpha H^2 + \beta \dot H)} \ln \left(\frac{H_{1}^{2}}{(\alpha H^2 + \beta \dot H)}\right) \Delta^{\prime}(x)\right\} 
\end{equation}
For extended Barrow entropy scenario, we have obtain the second FLRW equation by using the first law of thermodynamics as,

\begin{eqnarray}
\label{22}
(2 \alpha H \dot H + \beta \ddot H)
\left\{1+\Delta(x)+\frac{H_{1}^{2}}{\alpha H^2 + \beta \dot H} \ln \left(\frac{H_{1}^{2}}{\alpha H^2 + \beta \dot H}\right) \Delta^{\prime}(x)\right\} &\times&
\nonumber\\
\left(\frac{H_{1}^{2}}{\alpha H^2 + \beta \dot H}\right)^{\Delta(x)} &=&  -4 \pi G (p+\rho),  
\end{eqnarray}
where, $p$ and $\rho$ stand for the energy density and pressure of the matter, respectively.  As we have observed that, the FLRW equations is affected by the running behaviour of $\Delta(x)$ as compared with the constant exponent (see Eq. (12)). On integrating the above equation and using conservation law, the first FLRW equation can read as
\begin{equation}
\label{23}
-\left.H_{1}^{2}\left\{x^{-1+\Delta(x)}+2 \int^{x}  x^{-2+\Delta(x)} d x\right\}\right|_{x=\eta}=\frac{8 \pi G}{3} \rho+\frac{\Lambda}{3} .
\end{equation}
here, $\eta = \frac{H_{1}^{2}}{\alpha H^2 + \beta \dot H}$ and  $H \frac{d H}{d x}=-\frac{H_{1}^{2}}{2 x^{2}}$. The modified expression of FLRW equations with variable exponent in the context of the Barrow entropic energy can be seen in Eqs. (22) and (23).  Barrow entropic energy density  $\rho_{\mathrm{B}}$ and pressure $p_{\mathrm{B}}$ with variable exponent can be read as
\begin{equation}
\label{24}
\rho_{\mathrm{B}}=\frac{3}{8 \pi G}\left(H^{2}+\left.H_{1}^{2}\left\{x^{-1+\Delta(x)}+2 \int^{x} x^{-2+\Delta(x)} d x\right\}\right|_{x=\eta}\right),
\end{equation}
and
\begin{eqnarray}
 \label{25}
p_{\mathrm{B}} &=& -\rho_{\mathrm{B}} + \frac{ 1}{4 \pi G}
\bigg[(2 \alpha H \dot H + \beta \ddot H)
\nonumber\\
&\times&
\left\{1+\Delta(x)+\frac{H_{1}^{2}}{\alpha H^2 + \beta \dot H} \ln \left(\frac{H_{1}^{2}}{\alpha H^2 + \beta \dot H}\right) \Delta^{\prime}(x)
\left(\frac{H_{1}^{2}}{\alpha H^2 + \beta \dot H}\right)^{\Delta(x)}  \right\}
- \dot H \bigg]
 \end{eqnarray}
In order to obtain an explicit expression of energy density and pressure from the above equation and to integrate Eq. (24) analytically, a functional form of $\Delta$ is required. At the late time, if the exponent $\Delta(x)$ turns into constant, the outcome of the expended scenario will agree with the Barrow dark energy (BDE) model, where the entropy exponent is assumed to be constant.  If the exponent $\Delta$ is assumed in such a manner that at low and high energy scales the values of the exponent diverge from the regular value $1$. But for the transitional scales, the values remain close to unity. It results in a unified scenario with early inflation, late dark energy, and an intermediate deceleration phase. From Eqs. (24) and (25), we define the ``EoS parameter for barrow entropy" as
\begin{eqnarray}\label{26}
\omega_{\mathrm{B}} &=& -1+ \frac{2 }{3 }\nonumber\\
&\times& \frac{(2 \alpha H \dot H + \beta \ddot H)
		\left\{1+\Delta(x)+\frac{H_{1}^{2}}{\alpha H^2 + \beta \dot H} \ln \left(\frac{H_{1}^{2}}{\alpha H^2 + \beta \dot H}\right) \Delta^{\prime}(x)
	\left(\frac{H_{1}^{2}}{\alpha H^2 + \beta \dot H}\right)^{\Delta(x)}  \right\}
	- \dot H }{x^{\Delta(x)}+2 x \int^{x} x^{-2+\Delta(x)}dx+1}
\end{eqnarray}
For the extended Barrow entropy scenario, where the exponent varies with the cosmic evolution of the universe, the efficient EoS parameter can be seen in Eq. (26). Presuming the scenario, the correspondence between the generalized holographic energy density and the extended form of the Barrow energy density can be established. Nojiri-Odintsov \cite{ref35dd} have proposed the generalized cut-off for holographic dark energy (HDE). According to the holographic principle, the HDE energy density is inversely proportional to the square of the Generalized HDE cut-off $L_{GO}$, in particularly, $\rho_{\mathrm{hol}}=\frac{3 c^{2}}{\kappa^{2} L_{\mathrm{GO}}^{2}}$, where $\kappa^{2}=8 \pi G$, $G$ is the gravitational constant.  Here, we use two different cut-off, particle horizon  $L_{\mathrm{p}} \equiv a \int_{0}^{t} \frac{dt}{a}$ and the future event horizon $L_{\mathrm{f}} \equiv a \int_{t}^{\infty} \frac{dt}{a} $.\\
The Hubble parameter can be determined as $H(L_p, \dot L_p) = \frac{\dot L_p -1}{L_p}$ and  $ H(L_f, \dot L_f) = \frac{\dot L_f +1}{L_f}$.  Holographic cut-off (denoted by $L_B$) corresponding to the extended Barrow entropic scenario in terms of $L_p$ and its derivative is given as
\begin{eqnarray}\label{27}
\frac{3 c^{2}}{\kappa^{2} L_{\mathrm{B}}^{2}}&=&\frac{3}{8 \pi G}\Bigg[  \left. H_{1}^{2} 
\left\{x^{-1+\Delta(x)}+ \int^{x} 2 x^{-2+\Delta(x)} dx\right\} 
\right
|_{x=\frac{H_{1}^{2}}{{\alpha     \left(\frac{\dot{L}_{\mathrm{p}}}{L_{\mathrm{p}}}-\frac{1}{L_{\mathrm{p}}}\right)^{2} + \beta \left(\frac{\ddot{L_p}}{L_p } -\left(\frac{\dot{L_p}}{L_p}\right)^{2}+ \frac{\dot{L_p}}{L_p ^2}
			\right) }} }
 \nonumber\\
 + \bigg(\frac{\dot{L}_{\mathrm{p}}}{L_{\mathrm{p}}}-\frac{1}{L_{\mathrm{p}}}\bigg)^{2} \Bigg]
\end{eqnarray}
 
 in term of future event horizon $L_f$ and its derivative
 \begin{eqnarray}\label{28}
 \frac{3 c^{2}}{\kappa^{2} L_{\mathrm{B}}^{2}}&=&\frac{3}{8 \pi G}\Bigg[  \left. H_{1}^{2} 
 \left\{x^{-1+\Delta(x)}+2 \int^{x} x^{-2+\Delta(x)} dx \right\} 
 \right
 |_{x=\frac{H_{1}^{2}}{{\alpha     \left(\frac{\dot{L}_{\mathrm{f}}}{L_{\mathrm{f}}}-\frac{1}{L_{\mathrm{f}}}\right)^{2} + \beta \left(\frac{\ddot{L_f}}{L_f } -\left(\frac{\dot{L_f}}{L_f}\right)^{2}+ \frac{\dot{L_f}}{L_f ^2}
 			\right) }} }
 \nonumber\\
 + \bigg(\frac{\dot{L}_{\mathrm{f}}}{L_{\mathrm{f}}}-\frac{1}{L_{\mathrm{f}}}\bigg)^{2} \Bigg]
 \end{eqnarray}
 Along with the initial Friedmann equation, it is also required to establish the correspondence between the ``EoS parameters of the generalized HDE and BHDE models". So, we define the  EoS parameter corresponds to the cut-off $L_B$. It is equivalent to the energy density of HDE ${\rho^{(B)}_{hol}} = \frac{3 c^2}{{\kappa}^2 {L_B}^2}$. Following the conservation of $\rho_{hol}^B$, the EoS parameter $W_{hol}^B$ can be determined as:
 \begin{equation}
 \label{29}
 W_{\text {hol }}^{(\mathrm{B})}=-1+\left(\frac{2}{3 H L_{\mathrm{B}}}\right) \frac{d L_{\mathrm{B}}}{d t}
 \end{equation}
 
From Eqs. (26) and (29), the two EoS parameters $\omega_B$ and $W_{\mathrm{hol}}^{(\mathrm{B})}$ are found to be equivalent. 

 \section{Power law cosmology in Barrow entropy}
The type Ia supernovae observations \cite{ref1,ref2}, CMB anisotropies \cite{ref3} and recently Planck Collaborations \cite{ref6} have confirmed that the present Universe is in an accelerating phase. Therefore, to explain current accelerated expansion of the Universe, we assume scale factor in the from 
\begin{equation}
\label{30}
a(t) = k t^m
\end{equation}
where, $k>0$ is constant and $m>0$ is real which describes the development of scale factor in distinct eras of the evolution of the universe i..e. for $m = 1$ defines the marginal inflation $(a \propto t)$, $m = \frac{1}{2}$ for radiation dominated era, $m = \frac{2}{3}$ shows matter-dominated era and $m = \frac{4}{3}$ describe the accelerating era of the universe ($a \propto t^\frac{4}{3}$) \cite{ref45}. This form of a(t) describes the power law cosmology and resembles the late time acceleration of the universe. Power-law cosmology is an intriguing solution for dealing with some unusual challenges like flatness, horizon problem, etc. Kumar \cite{ref51} used power-law with $H(z)$ and SNe Ia data to analyze cosmological parameters. Rani et al.,\cite{ref51aa} also examined the power-law cosmology with statefinder analysis. Some important applications of power law cosmology are given in the References Kumar \cite{ref51} and Sharma et al. \cite{ref51a}.\\
 
According to cosmological observations in cosmology, the Hubble parameter $H$ and deceleration parameter $q$ are some of the most important observational quantities. These are defined as
\begin{equation}
\label{31}
H = \frac{\dot a}{a} = \frac{m}{t}
\end{equation}

\begin{equation}
\label{32}
q = \frac{- a \ddot a}{\dot a^2} = \frac{1}{m} - 1
\end{equation}
The relationship between redshift and scale factor is defined as $a = \frac{a_0}{1+z}$. In term of redshift z, the Hubble parameter is read as
\begin{equation}
\label{33}
H(z) = -\frac{1}{1+z} \frac{dz}{dt} 
\end{equation}
With the help of equation (13) and (16), we obtain the following expressions
\begin{equation}
\label{34}
H(z) = m \left(\frac{a_0}{k} \right)^{\frac{-1}{m}} (1+z)^{\frac{1}{m}} 
\end{equation}
The Hubble parameter in the term of redshift, expressed as $H(z) = H_0 (1+z)^{1/m}$, shows the expansion history of the universe in power law cosmology. Which depends on the model parameters $H_0$, $m$ and $k$ under consideration in view of observational $H(z)$ datasets in the redshift range $0 \leq z \leq 2.36$.\\

 \section{Observational constraints on model parameters}
Numerous authors estimated the Hubble constant in the range of 67 to 74 by utilizing the observational data of Hubble Telescope \cite{ref46}, ``Cepheid variable observations \cite{ref47, ref48}", ``WMAP seven-year data \cite{ref49}", and other sources \cite{ref50, ref51, ref52, ref53,ref54}. We take into account the latest 57 OHD data points. \\
 
To constrain the model parameters, we employ the Markov Chain Monte Carlo (MCMC) approach. By fitting the current model to the latest 57 OHD points in the redshift range $ 0 $ to $2.36$, based on the emcee python package, the model parameter $ m$, $k $ are estimated. The model parameters $ k = 65.4 \pm 1.1$, $ H_0 = 67.3 \pm 1.1$, and $ m = 1.0213 \pm 0.0071$ are found to be the best fits for the existing model to the $H(z)$ data at a $68\%$ CL. The current model's fitted value of $H_0 $ agrees well with that of the Plank collaboration.
 \begin{figure}[H]
 	\centering
 	\includegraphics[scale=0.7]{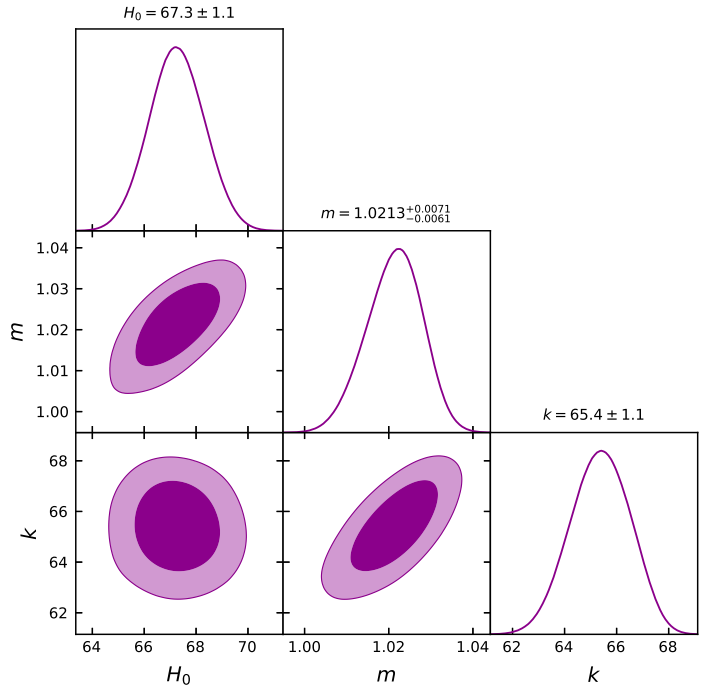}
 	\caption{The contour plots of the model parameters m, k, $H_0$ with
 		$1-\sigma$ and $2-\sigma$ confidence limits for 57 OHD points.}
 \end{figure}
Many authors have considered the value of the Barrow exponent $\Delta$ to be in the range $0\leq\Delta\leq1$ (see refs. \cite{ref41,ref41a,ref21,ref24,ref25,ref29}). Adhikary {\it et al.} \cite{ref30} recently took into account the value of $\Delta$ in the range $0\leq\Delta\leq 0.4$. Capozziello {\it et al.} \cite{ref54} recently discussed the Big Bang Nucleosynthesis (BBN) constraints on Barrow exponent $\Delta$, where $\Delta$ should be inside the bound $\Delta < 1.4\time 10^{-4}$ to spoil the BBN epoch, indicating that the deformation from standard Bekenstein-Hawking expression should be small as expected. By following the values of $\Delta$ in the range $0.45\leq\Delta\leq 0.95$, Mamon {\it et al.} \cite{ref21} detailed the dynamics of the BHDE model and noted that their model lies in the quintessence regime and phantom regime. Following the study given above, we have used the value of $\Delta$ in the range $0.05\leq\Delta\leq 0.25$ to characterize the dynamics of our model. We also intend to investigate the BBN using the Barrow entropy and spacetime thermodynamics, as stated in \cite{ref21}, within the context of changed cosmology.
Figure $1$ shows the contour plots of the model parameters $m$, $k$, $H_{0}$ with $1-\sigma$ and $2- \sigma$ confidence limits for 57 OHD points.
 \begin{figure}[H]
 	(a)\includegraphics[width=8cm,height=6cm,angle=0]{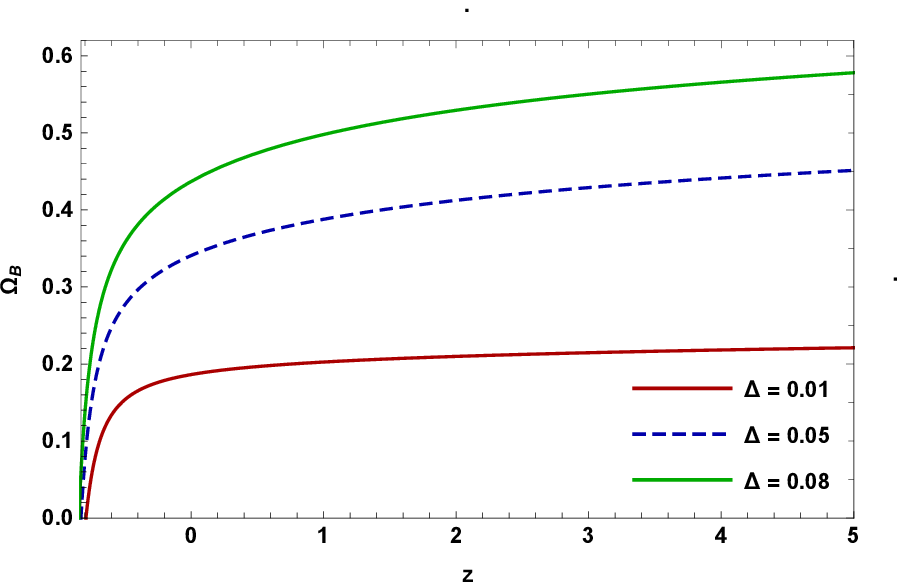}
 	(b)\includegraphics[width=8cm,height=6cm,angle=0]{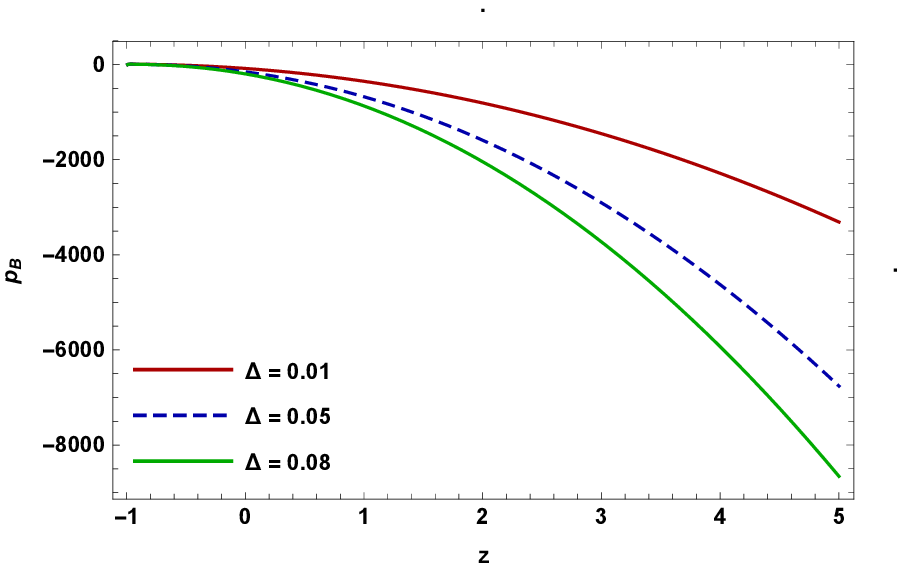}
 	\caption{(a)  Plot of density parameter $ \Omega_B$, (b) Plot of pressure for BHDE. }
 \end{figure}
Fig. 2(a) depicts the nature of density parameter $\Omega_{B}$ of BHDE against redshift $z$. The energy density parameter has positive behavior as clear from the figure. Figure 2(b) shows the behavior of pressure $p_D$ for estimated values of model parameters $m= 1.0213, k = 65.4, H_0 = 67.3$. The `pressure' is always negative through the `entire evolution of the universe' as clearly seen from the figure.  
\begin{figure}[H]
 	\centering
 	\includegraphics[width=8cm,height=6cm,angle=0]{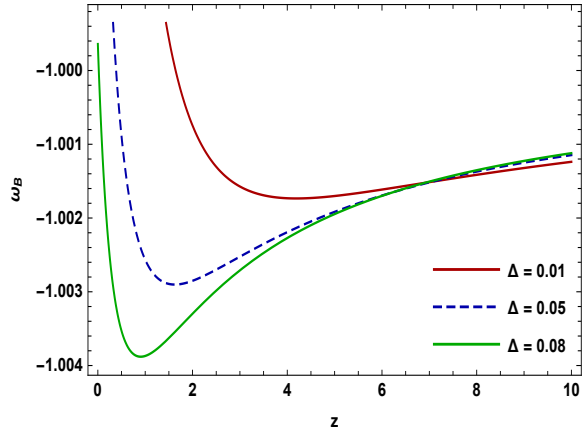}
 	\caption{Plot of EoS parameter for $m = 1.0213$ and $k = 65.4$. }
 \end{figure}

Figure $3$, exhibit the behavior of EoS parameter $\omega_{B}$ for the estimated values $m = 1.0213$ and $k = 65.4$. From figure it is clear that initially, the model lies in quitessence region crosses the $\Lambda$CDM and lies in  phantom region ($\omega_{B} <{-1} $) at late time. 
     
\section{Holographic quintessence model}
Various DE models have been investigated in the context of quintessence field. For dark energy, a number of different models have been proposed. Only distance measurements may be used to discriminate between two dark energy models that have the same assessment of the scale factor, and these tests are unable to distinguish between them. Because of this, it is critical to analyse the growth rate of disturbances in matter using the same scale factor for various models of dark energy in order to compare the results. Energy density and pressure for the quintessence scalar field model are given by \cite{ref55}
   
\begin{equation}
   \label{35}
   \rho_B = \frac{\dot \phi^2}{2} + V(\phi); ~~ p_B = \frac{\dot \phi^2}{2} - V(\phi)
\end{equation}
from the above equation, we obtain
   \begin{equation}
   \label{36}
   {\dot \phi^2} = \rho_B + p_B, ~~ V(\phi) = \frac{\rho_B - p_B}{2} = 
   \frac{(1- \omega_B)}{2} \rho_B
   \end{equation}
   
 For an accelerated expansion we need flat potential, which is obtained by the condition $\dot\phi^{2} <V$.
The range of EoS parameter for quintessence scalar field $\phi$ lies in the region $(-1\leq\omega\leq 1)$  where $\omega_{D}=-1$ relates to the condition of the slow-roll
limit  $\dot\phi^{2} \leq V$. Condition  $\dot\phi^{2} \geq V(\phi)$ denotes the presence of the stiff matter in the Universe. Due to some type of phantom  dark energy, the area where the equation of state $\omega_{D}\leq -1$  for  $\dot\phi^{2} < V(\phi)$  is often referred to as \cite{ref56}. The equation for the scalar field and scalar potential is found by
the equation (35) and (37) as
\begin{equation}
  \label{37}
   {\dot \phi^2} = 
    \frac{1}{4 \pi G} \left[ \frac{({1+ \Delta})}{H} (2 \alpha H \dot H + \beta \ddot H) \left(\frac{{{H_1}^2}}{ \alpha H^2 + \beta \dot H} \right)^{\Delta} - \dot H   \right]
\end{equation}

\[
 {V (\phi)} = \frac{3}{8 \pi G} \left( H^2 - 
 \frac{({1+ \Delta})}{{(1- \Delta)}}
 {H_1}^2  \left(\frac{{{H_1}^2}}{ \alpha H^2 + \beta \dot H} \right)^{\Delta - 1} \right)
\]
  \begin{equation}
  \label{38}
  + \frac{1}{8 \pi G} \left[ \frac{({1+ \Delta})}{H} (2 \alpha H \dot H + \beta \ddot H) \left(\frac{{{H_1}^2}}{ \alpha H^2 + \beta \dot H} \right)^{\Delta} - \dot H   \right]
  \end{equation} 
\begin{figure}[H]
	(a)\includegraphics[scale=0.33]{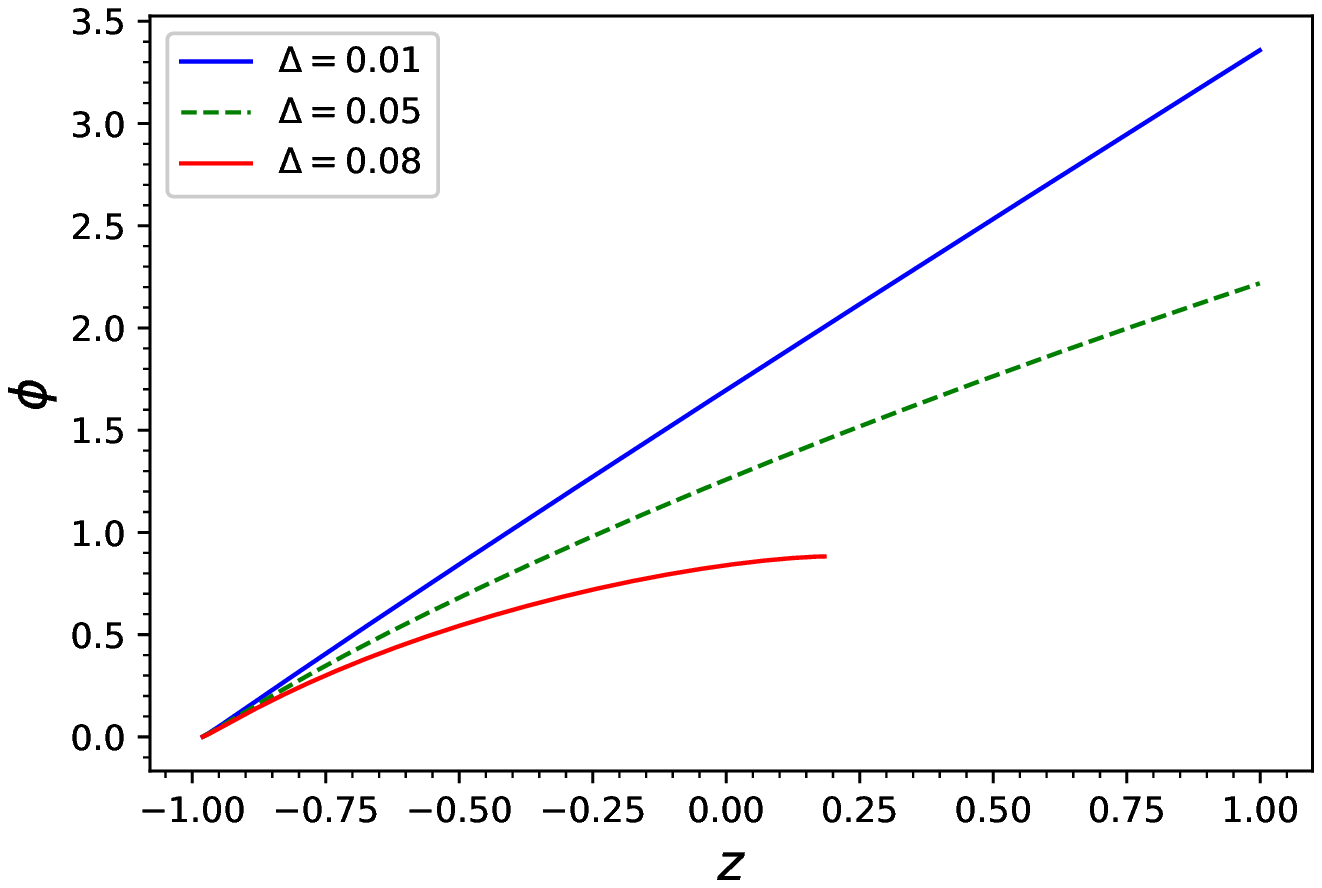}
    (b)\includegraphics[scale=0.55]{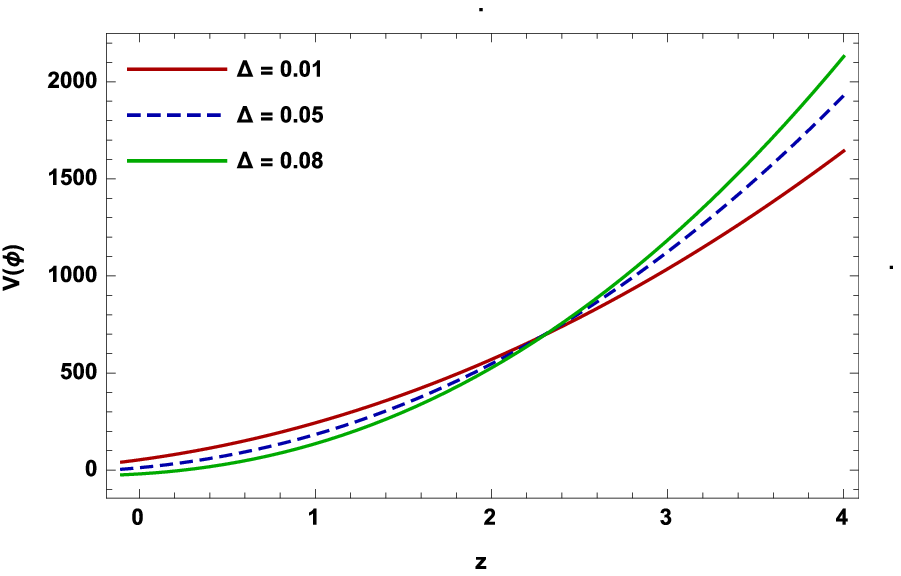}
    (b)\includegraphics[scale=0.33]{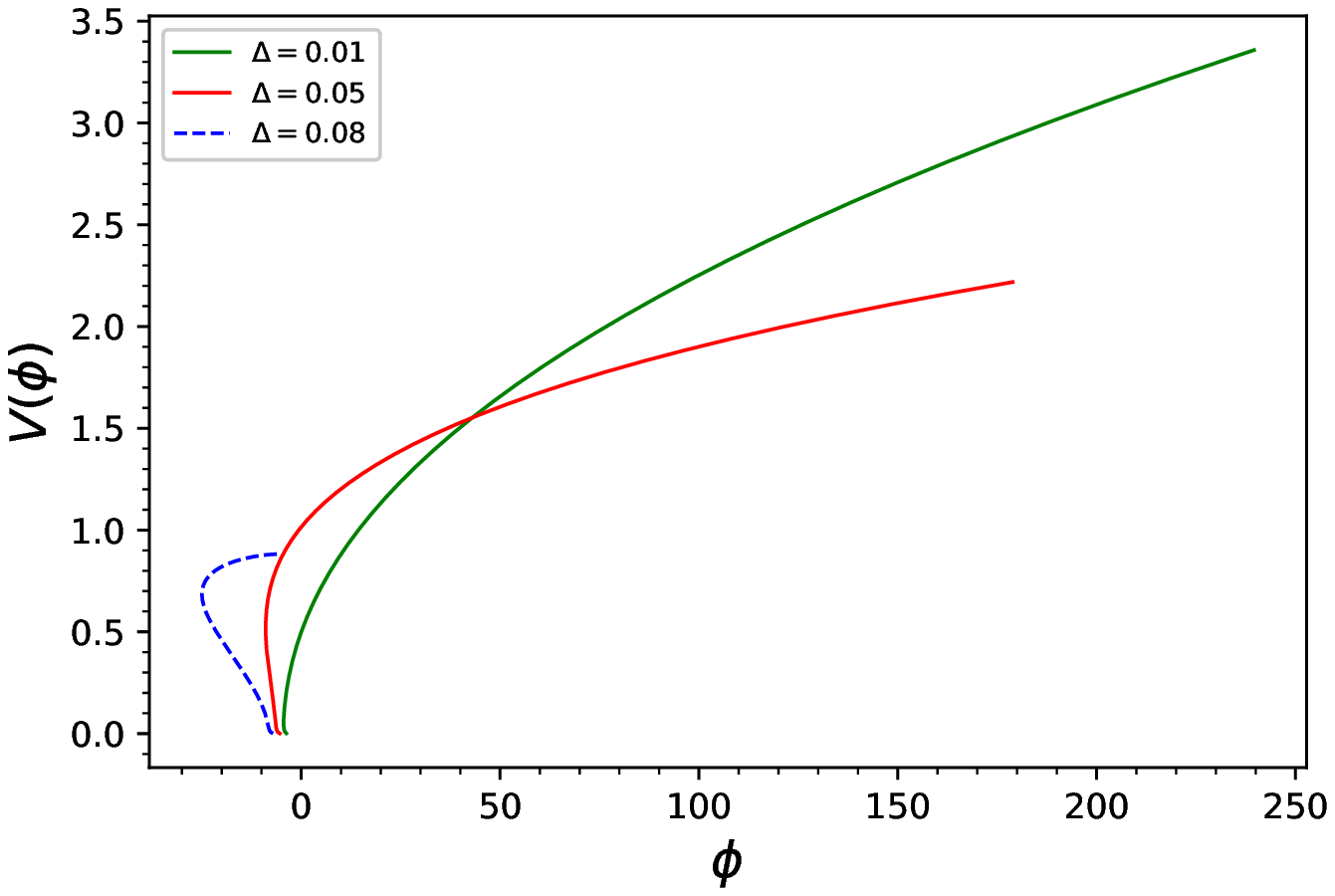}
	\caption{(a) Figure of $ \phi$ vs $z$,  (b)  Figure of $V(\phi)$ versus  $z$  (c) Figure of $ V(\phi)$ vs $\phi$}
\end{figure}  
Figure 4(a) and 4(b) depicts the evolution of scalar field $\phi$ and potential $V(\phi)$ of quintessence model with respect to redshift z. It has been plotted for the estimated values of parameters $m, k, H_0 $ 
( $m= 1.0213, k = 65.4, H_0 = 67.3$). 
For these suitable choices of the parameters the field gets trapped in the local minimum because the kinetic energy during a scaling regime is small. The field then enters a regime of damped oscillations leading to an accelerating universe.

\section{Holographic dilation field}

A dilaton scalar field, originated from the lower-energy limit of string theory \cite{ref57}, can also be assumed as a source of DE. This model appears from a four-dimensional effective low-energy string action \cite{ref57} and includes higher-order kinetic corrections to the tree-level action in low energy effective string theory. The coefficient of the kinematic term of the dilaton can be negative in the Einstein frame, which means that the dilaton behaves as a phantom-like scalar field. Energy density and pressure (Lagrangian) of the dilaton DE model are given, by \cite{ref59} 

\begin{equation}
\label{39}
\rho_{B}=-X+3 c e^{\lambda \phi} X^{2} = - X +3 f(\phi) X^2
\end{equation}
\begin{equation}
\label{40}
p_{B}=-X+c e^{\lambda \phi} X^{2} = - X +  f(\phi) X^2
\end{equation}
where $c$ is a positive constant and $X= \frac{\dot{\phi}^{2}}{2}$.
Equation of state parameter $\omega_B$ for the dilaton
scalar field can be obtained from  

\begin{equation}
\label{41}
\omega_{B}=\frac{-1+c e^{\lambda \phi} X}{-1+3 c e^{\lambda \phi} X} 
\end{equation}
From the above equation we find the value of X, 

\begin{equation}
\label{42}
X = \frac{\omega_{B} -1}{(3 \omega_{B} - 1) c e^{\lambda \phi}}
\end{equation}
The scalar field is read as
\begin{equation}
\label{43}
\dot \phi^2 = \dfrac{c e^{\lambda \phi}}{2} \frac{(3 \omega_{B} -1)}
{( \omega_{B} - 1) } = (\rho - 3p )
\end{equation}

\begin{equation}
\label{44}
f {(\phi)} =  \frac{(\rho - p )}{2 X^2}
\end{equation}

\begin{figure}[H]
	(a)\includegraphics[width=8cm,height=6cm,angle=0]{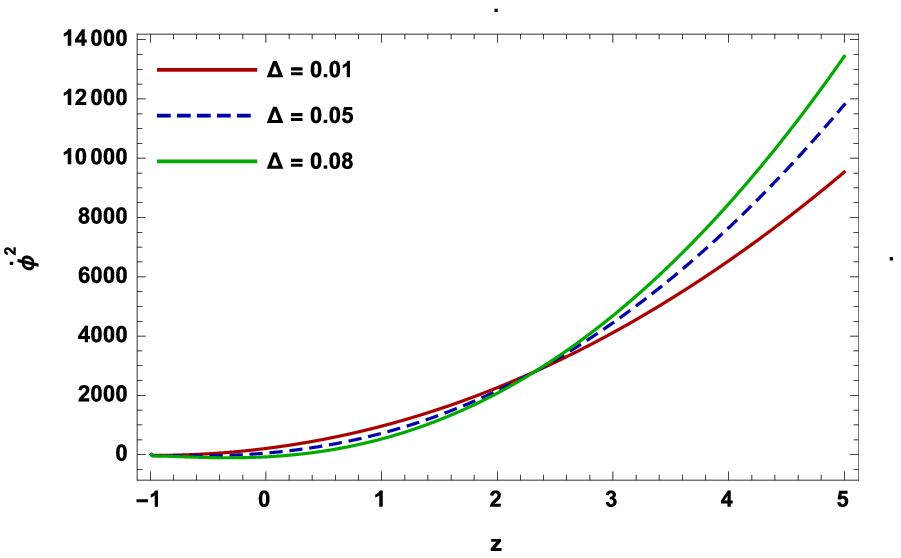}
	~~~~~~~~~~~~~~(b)\includegraphics[width=8cm,height=6cm,angle=0]{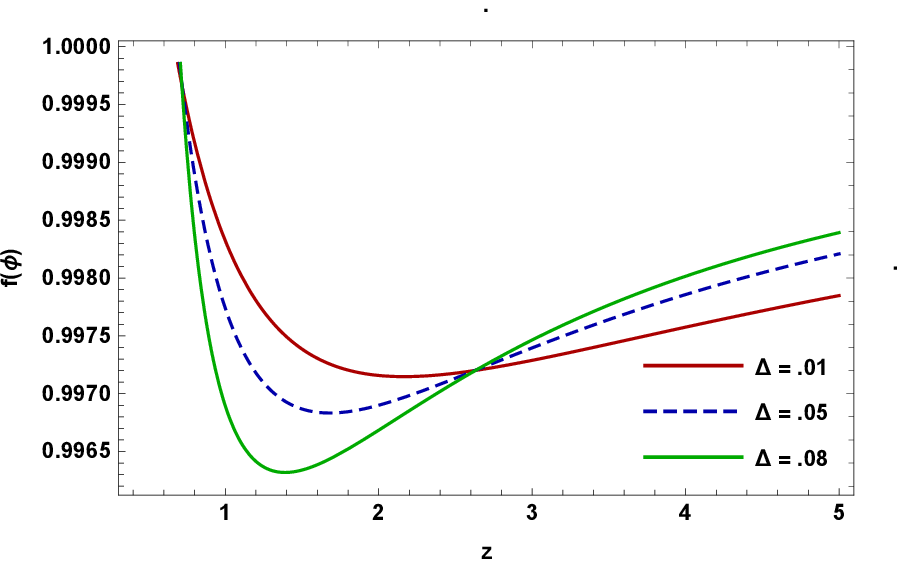}
	~~\caption{(a) Figure of $\dot \phi^2$ vs  $z$ , (b)  Figure of $f(\phi)$ vs  $z$ }
\end{figure} 
Figure 5(a) and 5(b) shows the variation of KE with variation of redshift $z$ for best fit values $m= 1.0213, k = 65.4, H_0 = 67.3$. It has been ploted for three different values of Barrow exponent choosing $\Delta = 0.01 ~ $, $\Delta = 0.05 ~ $\& $\Delta = 0.08 ~ $. From the figure, we observe that scalar field $\phi(z)$ rises as z increases.

\section{Concluding remarks}
In this paper, the Authors have described the dynamics of the universe via graphical representation by assuming the scale factor as $a = k t^m$. The value of free parameters $k, m$ are estimated on 57 OHD points utilizing MCMC (Markov Chain Carlo) method is used. Nojiri et al. \cite{ref35} proposed the extension of the generalized BHDE model. In this manuscript, the authors have revisited the extension of BHDE and its equivalence with generalized holographic dark energy adopting ``Generalized HDE cut-off \cite{ref35dd} for the model". We have also explained the ``dynamics of quintessence and dilation scalar field models".
 
\begin{itemize}
\item Figure 1 demonstrates the ``2-Dimensional contour plots and 1-Dimensional marginal plots". The best fitted values of the model parameter are determined to be $m = 1.0213, k = 65.4 H_0 = 67.3 $.

\item From Fig. 2, it is clear that the `energy density ($\rho_{B}$)' is positive and `cosmic pressure ($p_{B}$)' is negative through the evolution of the universe for BHDE with Generalized HDE cutoff \cite{ref35dd}.

\item Figure 3, describes the behavior of EoS parameter $\omega_{B}$ for the BHDE model in the reference of the Generalized HDE cut-off. The EoS parameter lies in the phantom era ($\omega_D \leq{-1}  $) and remains negative in the entire evolution of the universe.
	
\item Figs. 4 $\&$ 5 depict the quintessence and dilation of the model. The model has an ``attractor solution with accelerated expansion, and it also depends on the field's inverse square" to be able to achieve the dilation holographic correspondence. From Fig 4, we notice that ``$\dot{\phi}^{2} < V(\phi)$". The potential for the quintessence model is decreasing the function, which indicates to an accelerated expansion of the universe. Similarly, ``for dilation model, $\dot{\phi}^{2} < f(\phi)$ and $f(\phi)$ is also a decreasing function".
	
The solution suggested in this study may therefore be helpful in better understanding the generalization of HDE theories in the history of the cosmos.
	
\end{itemize}
\section*{Acknowledgement}
A. Pradhan thanks the IUCAA, Pune, India, for providing support and facility under the associateship program. The authors also express their gratitude to the reviewers for valuable comments and suggestions.

\end{document}